\documentclass{corr-aadebug}

\usepackage{alltt}
\usepackage{moreverb}
\usepackage{graphicx}

\renewcommand{\L}[1]{{\sffamily #1}}

\def\varset{{\cal V}}
\def\consset{{\cal C}} 
\def\saut{\\[7pt]}
\def\et{,\ \ }
\def\reduce{\L{reduce}}
\def\reject{\L{reject}}
\def\solution{\L{solution}}
\def\newConstraint{\L{newConstraint}}
\def\newVariable{\L{newVariable}}
\def\awake{\L{awake}}
\def\suspend{\L{suspend}}

\newcommand{\reglepropa}[4]{\makebox[2.5cm][r]{{\bfseries #1}}~$\displaystyle\frac{#2}{#3}~ #4$}

\newcommand{\Var}[1]{\mathbf{var}(#1)}
\newcommand{\Code}[1]{{\tt #1}}

\begin{document}
\corr{0309027}{171}
\runningheads{M. Ducass\'e et al.}{AADEBUG2003: Rigorous design of tracers}

\title{Rigorous~design~of~tracers: \\an experiment for
  constraint logic programming}

\author{Mireille~Ducass\'{e}\addressnum{1}, Ludovic~Langevine\addressnum{2}, Pierre~Deransart\addressnum{2}}

\address{1}{
IRISA/INSA de Rennes, Campus Universitaire de Beaulieu, 35042 Rennes Cedex, France
}

\address{2}{
INRIA Rocquencourt, BP 105, 78153 Le Chesnay Cedex, France}

\extra{1}{E-mail: Mireille.Ducasse@irisa.fr, \{Ludovic.Langevine, Pierre.Deransart\}@inria.fr}
\extra{2}{This work is partially supported
  by the French RNTL project OADymPPaC, Tools for dynamic
analysis of constraint logic programs, \href{http://contraintes.inria.fr/OADymPPaC/}{http://contraintes.inria.fr/OADymPPaC/}}

\pdfinfo{
/Title (AADEBUG2003: Rigorous design of tracers)
/Author (Mireille Ducass\'e et al.)
}

\begin{abstract}
In order to design and implement tracers, one must decide {\em what}
exactly to trace and {\em how} to produce this trace.
On the one hand, trace designs are too often guided by implementation concerns and are
not as useful as they should be.
On the other hand, an interesting trace which cannot be produced
efficiently, is not very useful either.
In this article we propose a methodology which helps to efficiently
produce accurate traces.
Firstly, design a formal specification of the trace model.
Secondly, derive a prototype tracer from this specification.
Thirdly, analyze the produced traces.
Fourthly, implement an efficient tracer.
Lastly, compare the traces of the two tracers.
At each step, problems can be found. In that case one has to iterate
the process.
We have successfully applied the proposed methodology to the design
and implementation of a real tracer for constraint logic programming
which is able to efficiently generate information required to build
interesting graphical views of executions.
\end{abstract}


\section{Introduction}

Designing and implementing tracers is a difficult task. 
One must decide {\em what} exactly to trace and {\em how}
to produce this trace. 
On the one hand, trace designs are too often guided by implementation
concerns and are not as useful as they should be.
On the other hand, an interesting trace which cannot be produced
efficiently, is not very useful either.

Some sort of instrumentation is required to produce the trace
information. This instrumentation can be done at different levels, for
example the user programs can be transformed at source or compiled
levels. Another possibility is to plant trace hooks in the language
interpreter or emulator. All these possibilities have their advantages
and drawbacks, deciding for one is not straightforward. For example,
source level transformation does not require to open the compiler
sources and can be efficient enough~\cite{ducasse2000}. Instrumenting
an interpreter is in general easier than working directly in the
compiler and the loss in performance might be acceptable.

However, in order to have at the same time a faithful and efficient
enough tracer, people often decide to instrument the language emulator
or the user compiled code. In both cases, they have to dive deeply
into non-obvious code. This is a tricky task, especially if the tracer
developers have not been involved in the development of the
compiler. At this stage, it is necessary that the trace model is
stable. Instrumenting at low level requires a lot of tedious work, it
is essential to know exactly what is expected before starting.

The first stage of the design, namely deciding what to trace, is,
therefore, critical.
However, deciding what to trace out of the blue is not obvious. It is
often when people see the output of a tracer that they can tell
whether the information is relevant and helpful.

To solve this apparent contradiction, we have conceived a methodology
that we have successfully applied to the design and implementation of
a real tracer:
\begin{enumerate}
\item Design a formal specification of the trace model based on an
  abstraction of the operational semantics of the language.
\item Derive a prototype tracer from this specification.
\item Analyze the produced traces to update or validate the trace
  model. This analysis can be partially automated.
\item Implement an efficient tracer.
\item Validate the efficient implementation using the prototype.
\end{enumerate}
At each step, problems in the model or in implementations can be
found. In that case one has to iterate the process.\\

The advantages of the approach are as follows.
Firstly, the formal specification helps to produce a trace model which
gives an accurate picture of the executions. When examining existing
tracers, one too often has the feeling that they produce whatever
information is easily available, hoping that the users will manage
with the holes and the noise. While users often manage, mostly because
they have no other choice, we claim that tracers with clean semantics
are much more helpful.
Secondly, in our approach the prototype is systematically derived from
the trace model. It is therefore easy to produce and the resulting
traces are faithful to the model. The prototype can easily produce
trace samples.
Thirdly, analyzing the trace samples enables people to tune the trace
model. When an automated trace analyzer is used, this analysis can be
systematic and thorough.
Fourthly, when the tracer developers reach the stage where they have
to implement a low-level tracer, they know which information is crucial
and which one can be escaped. If implementation compromises have to be
made, people have rational arguments to take their decisions.
Lastly, comparing the actual output of the low-level implementation
against the expected trace produced by the prototype is a good way to
validate the quality of the implementation.\\

Note that the execution of the prototype tracer can afford to be
slow. Indeed, it is used only at development time for qualitative
reasoning, firstly to tune the trace model, then to test the
implementation.  Furthermore it is meant to be used on small programs
only.  The only requirement of these programs is that they altogether
exhibit all the characteristics of the traced language.  In order to
test the performance of the final tracer, big programs have, of
course, to be traced but with the low-level implementation not with
the prototype.\\

The first three steps of this methodology had been applied to the
{\em retrospective} design of a Prolog tracer \cite{jahier01b}. 
The formal specification step enabled to specify a variety of existing
trace models for Prolog which were shown to be minor variants of each
others.  The result was a unified view of numerous models which were
initially proposed without much rationale to support them.
Furthermore a number of implementation issues regarding variable
identifiers were detected.

More interestingly, we have used all the steps of the methodology to
design from scratch a tracer for constraint logic programming over
finite domains(CLP(FD) in the following). A detailed description of
this tracer can be found in~\cite{langevine03}.
The low-level tracer has been implemented inside the compiler of
Gnu-Prolog~\cite{gnuprolog} by somebody who did not previously know
the implementation of the compiler.
This experiment has been done within a project on visualization of
constraint program executions, with academic and industrial partners
working on different platforms. Both the formal specification of the
trace model and the prototyping capabilities have enabled us to
discuss with all the partners of the project and to make sure that the
designed trace is indeed matching the needs of everybody. 
Some partners build graphical views \cite{simonis2000}, other provide
explanations of failure for over-constrained
problems~\cite{jussien00}. They all work with parts of the specified
trace.
As the design of the trace model was not connected to a particular
implementation we could design a model which is generic enough to be
specialized at low cost for two different styles of constraint
satisfaction (CSP~\cite{laburthe01} and CLP~\cite{APT00}).
Furthermore, some parts of the designed information, for example the
list of variables and the constraint identifiers, have been proved
essential for the quality of the views but they are not obvious to
gather from the implementation in the emulator. Starting from the
formal specification and knowing that they were absolutely needed,
made them not too difficult to implement. One can conjecture that, had
we started to implement straightaway the low-level implementation,
these kinds of information would probably have never been provided by
the tracer. Not surprisingly, in the tracers of the CLP systems,
Sicstus Prolog~\cite{agren02} and Ilog Solver~\cite{IlogSolver01}, in
order to get the list of variables, one has to traverse a huge part of
the trace. This is especially tedious to do for users.\\

The contribution of this article is to propose a methodology to
rigorously design and validate tracer implementations.  To our best
knowledge, this had not been done before. This methodology has been
successfully tested against the design and implementation of a real
tracer.\\

The next sections present the methodology in more details and
illustrate it with samples of the experiment done with the design,
implementation and validation of a tracer for Gnu-Prolog.  The
explanations are meant for readers with no previous knowledge of
constraint solving. 
Section~\ref{sec:form-spec-trace} presents the trace
model formalization step.
Section~\ref{sec:impl-prot-trac} describes how to systematically derive an
instrumented meta-interpreter from the formal specification.
Section~\ref{sec:trace-analysis} outlines the trace analysis step. In
particular, it briefly presents the connection to a trace analyzer and
lists some interesting graphical views for CLP(FD).
Section~\ref{sec:effic-impl} discusses the 
low-level implementation of tracers.
Section~\ref{sec:validation} sketches the validation of the
efficient tracer against the prototype implementation.

\section{Formal specification of a trace model}
\label{sec:form-spec-trace}

The first step of our methodology is to specify a trace model, namely
what information about program executions should be given.  In our
sequential language context, a trace is a sequence of events. An event
represents an interesting execution step, it can be seen as an
execution breakpoint to which information is attached.

Most specifications of trace models are informal, when they exist at
all. However, an informal specification is prone to misinterpretation
by both developers and users. Indeed, in order to denote the events of
interest, one has first to be able to denote events, and this is not
easy in an informal way.

In the following, we describe two ways that have been tried to provide
a formal specification of trace model, firstly an existing operational
semantics has been instrumented and secondly an abstract operational
semantics has been specified. We then give some details of the second
experiment.

\subsection{Instrumenting an existing operational semantics}

Some programming languages have precise operational semantics which
rigorously specify the computation steps. Hence the computation events
are clearly denoted. In such a case, it is relatively easy to formally
specify a trace model as shown by Jahier et al. with the Prolog tracer
retro-speci\-fica\-tion~\cite{jahier01b}.  There were already numerous
operational semantics available for Prolog. The continuation passing
semantics of Nicholson and Foo \cite{nicholson89} was used. Then
instrumentations inside the operational semantics were specified.
Only 2 rules of one line each had to be instrumented in order to get
the ``standard'' trace of Prolog. The instrumentation itself is slightly
tricky but it can be understood without ambiguity when examining a
formal specification of 10 lines.
The detailed description of this experiment is out of the scope of
this article. Indeed, the explanations of operational semantics
requires a couple of pages, understanding them requires a good
knowledge of Prolog and this article is not aiming at Prolog
specialists.

\subsection{Specifying an abstract operational semantics}

An operational semantics is not always available, for example we are
not aware of any for the C language. In such a case, starting the
design of the tracer by designing a complete operational semantics is
certainly an overkill.
An operational semantics specifies in detail the execution of a
program. From an operational semantics one can derive an
implementation of a compiler.  In order to design a tracer, less
information is usually needed than to implement a compiler. In that
case, an abstract operational semantics is sufficient.
The information given in an abstract operational semantics is correct
but not complete. It tells tracer developers what information should
be provided to users and it tells users how to interpret this
trace information.
In the case of CLP(FD) there was no operational semantics that we
could use to specify a tracer and we designed an abstract operational
semantics as a set of state transition rules. .

\subsection{Informal presentation of domain reduction}

\begin{figure}[t]
  \begin{center}
    \includegraphics[width=14.5cm]{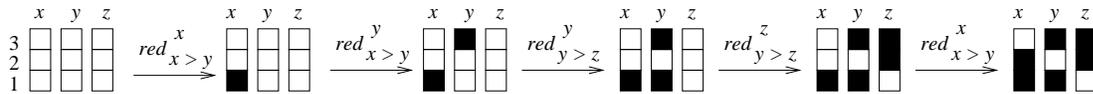}
    \caption{Application of reductions to the system $\{x> y; y >z\}$.}
    \label{FIG-REDUCTION}
  \end{center}
\end{figure}

Before we give examples of a formal specification of events we have to
informally explain how variable domains are reduced. This is an
essential mechanism of constraint propagation in the case of finite
domains. 
A CLP(FD) program searches a solution for a set of variables which take
values over finite domains and which must verify a set of constraints.
The evolution of the domains can be viewed as a sequence of
applications of reduction operators attached to the constraints.  Each
operator can be applied several times until the computation reaches a
fix-point~\cite{ferrand00}. This fix-point is the set of final domain
states.
An example of computation with reduction operators is shown in
Figure~\ref{FIG-REDUCTION}. There are three variables $x$, $y$ and $z$
and two constraints, $x > y$ and $y > z$.  A set of possible values is
associated to each variable. This set is called the domain of the
variable. The domain reduction consists in elementary steps that
remove inconsistent values from those domains. At the beginning, the
domain of $x$, $D_x$, the domain of $y$, $D_y$, and the domain of $z$,
$D_z$, are all equal to $\{1, 2, 3\}$. This is represented by three
columns of white squares.
Considering the first constraint, it appears that $x$ cannot take the
value ``1'', because otherwise there would be no value for $y$ such
that $x > y$; this inconsistent value is withdrawn from $D_x$. This
withdrawal is marked with a black square. 
In the same way, the value 3 is withdrawn from the domain of
$y$. 
Then, considering the constraint $y>z$, the sets $\{1\}$ and $\{2,
3\}$ are respectively withdrawn from $D_y$ and $D_z$. 
Finally, using again the first constraint to propagate the previous
reduction of $D_y$, $D_x$ is reduced to the singleton $\{3\}$.  The
fix-point is reached. The final solution is $\{x = 3, y = 2, z = 1\}$.


\subsection{Samples of formal event rules for CLP(FD)}
\label{formal:spec:section}

\begin{figure}
\noindent
\reglepropa{\newVariable{}}{x \not\in \varset}
{\phantom{xx,} \varset \gets \varset \cup \{x\} \et {\cal D} \gets {\cal D} \cup
  \{(x, D_x)\} \phantom{xx,}}
{\{D_x$ : initial domain of $x\}}
\saut
\saut
\reglepropa{\newConstraint{}}
{\phantom{xx,} c \not\in \consset \; \wedge \; \Var{c} \subset \varset
  \phantom{xx,}}
{\consset \gets \consset \cup \{c\}}{} 
\saut
\saut
\reglepropa{\reject}{\phantom{xx,} A = \{c\}  \; \wedge \; unsatisfiable(c,{\cal D}) \;
  \wedge \; R = \emptyset \phantom{xx,}}
{A \gets \emptyset \et R \gets \{c\}}{}
\saut
\saut
\reglepropa{\suspend}
{\phantom{xx,} A = \{c\}  \; \wedge \; no\_reduction(c,{\cal D})
 \; \wedge \; R = \emptyset \phantom{xx,}}
{A \gets \emptyset \et S \gets S \cup \{c\}}
{}
\saut
\saut
\reglepropa{\awake}{\phantom{xx,,}A = \emptyset{} \; \wedge \; wake\_condition(c)\;
  \wedge \; R = \emptyset \phantom{xx,,}}
{A \gets \{ c \}}{}
\saut
\saut
\reglepropa{\reduce}
{\; A = \{c\} \; \wedge \; x \in \mathrm{var}(c) \; \wedge \; W^c_x({\cal D})
  \neq \emptyset \; \wedge \; R = \emptyset \; }
{D_x \gets D_x - W^c_x({\cal D})}
{\left\{
\begin{array}{l}
W^c_x({\cal D}):
\hbox{ inconsistent values}\\
\hbox{  of }
x
\hbox{ for }
c
\hbox{ wrt }
{\cal D}
\end{array}\right\}}
\saut
\caption{Six rules of the abstract operational semantics defining six
  event types}
\label{event:rules:figure}
\end{figure}

Our abstract operational semantics is defined by a set of
transition rules between {\em observed states}.
An observed  state is a tuple containing in particular:
${\cal C}$, the set of constraints declared until this
state;\quad
${\cal V}$, the set of finite-domain variables declared until this
state; \quad
${\cal D}$, the set of domains of the variables declared until this state;
$A$, the set of active constraints ;\quad
$S$, the set of sleeping constraints;\quad
and
$R$, the set of rejected constraints which contains
unsatisfiable constraints.\\

Figure~\ref{event:rules:figure} gives six examples of transition rules
specifying types of events of interest. The complete trace model
contains thirteen event types.
Rule \newVariable{} specifies that a new  variable $x$ is introduced
in $\cal V$ and that its initial domain is $D_x$.
Rule \newConstraint{} specifies that the solver
introduces a new constraint $c$ in $\cal C$,  all
variables involved in $c$ are already defined. 
Rule \reject{} specifies that if the active constraint ($A = \{c\}$)
is unsatisfiable in the current state of the domains, it is put in the
set of rejected constraints, $R$. A constraint is unsatisfiable for
example when one of its variables has an empty domain.
Rule \suspend{} specifies that when an active constraint cannot reduce
any domain at the moment, it is suspended in $S$.
Rule \awake{} specifies that when the set of active constraints is
empty and some specific condition is fulfilled a suspended constraint
can be awoken and become active \footnote{Awakening conditions are
solver dependent, and usually contain the ``added value'' of each
solver. Some are therefore rather reluctant to show this condition in
the trace.}.
Rule \reduce{} specifies that if the active constraint, $c$, has a
variable, $x$, with inconsistent values in its domain, $W_x^c({\cal
D})$, these values are withdrawn from its domain $D_x$.

\subsection{A Trace Example}
\label{sec:trace-example}

\begin{figure}
\begin{alltt}\tt
 1 newVariable   v1 =[0-268435455]
 2 newVariable   v2 =[0-268435455]
 3 newConstraint c1  fd_element([v1,[2,5,7],v2])
 4 reduce   c1   v1 =[1,2,3]   W=[0,4-268435455]
 5 reduce   c1   v2 =[2,5,7]   W=[0-1,3-4,6,8-268435455]
 6 suspend  c1
 7 newConstraint c4 x_eq_y([v2,v1])
 8 reduce   c4   v2 =[2]       W=[5,7]
 9 reduce   c4   v1 =[2]       W=[1,3]
10 suspend  c4
11 awake    c1
12 reject   c1
...
\end{alltt}
\caption{A portion of Trace for\quad {\tt fd\_element(I,[2,5,7],A),
(A\#=I ; A\#=2)}.\quad $a$-$b$ means \emph{from $a$ to $b$} and
$a$,$b$ means \emph{$a$ and $b$}}
\label{fig:trace:example}
\end{figure}

Figure~\ref{fig:trace:example} presents the beginning of a trace of a
toy program in order to illustrate the event types described above.
This program, {\tt fd\_element(I, [2,5,7],A), (A\#=I ; A\#=2)},
specifies that {\tt A} is a finite domain variable which is in
$\{2,5,7\}$ and {\tt I} is the index of the value of {\tt A} in this
list; moreover {\tt A} is either equal to {\tt I} or equal to 2. The
first option is infeasible;  the trace shows the events related to the
failing attempt to satisfy it.

The trace can be read as follows. 
The first two events are related to the introduction of two variables
{\tt\small v}$1$ and {\tt\small v}$2$, corresponding respectively
to{\tt I} and {\tt A}.  In Gnu-Prolog variables are always created
with the maximum domain (from 0 to 268.435.455).
Then the first constraint is created: {\tt fd\_element} (event \#3).
This constraint makes two domain reductions (events \#4 and \#5): the
domain of the first variable ({\tt I}) becomes $\{1,2,3\}$ and the
domain of {\tt A} becomes $\{2,5,7\}$. 
After these reductions, the constraint is suspended (event \#6).
The next constraint, {\tt A\#=I}, is added (event \#7).
Two reductions are done on variables {\tt A} and {\tt I}, the only
possible value for {\tt A} and {\tt I} to be equal is {\tt 2} (events
\#8 and \#9).
After these reductions, the constraint is suspended (event \#10).
The first constraint is awoken (event \#11).
If {\tt A} and {\tt I} are both equal to {\tt 2}, {\tt I} cannot be
the rank of {\tt A}. Indeed, the rank of {\tt 2} is {\tt 1} and the
value at rank {\tt 2} is {\tt 5}.  The constraint is therefore
rejected (event \#12).
The execution continues and find the solution {\tt A=2} and {\tt I=1}.
This requires 20 other events not shown here

\subsection{Discussion}

Our semantics does not specify how the rules are applied but what
events are of interest and what information is available at each
event.
As already mentioned, it helps tracer developers to design the tracer
and it helps tracer users to interpret the produced traces.

For example, rule \reject{} tells the implementor that whenever the
solver finds an unsatisfiable constraint, the tracer should be called
with this constraint. The same rule tells users that when they see a
\reject{} event in a trace, the corresponding constraint has become
unsatisfiable and there was previously no rejected constraint.

Similarly, rule \reduce{} tells the implementor that, when the solver
processes a reduction, the tracer should be called with information
about the related constraint, the related variable and the new domain
$D_x$. The same rule tells users that when they see a \reduce{} event
in a trace, the reduction has been achieved on one of the variables of
the specified constraint, that there were inconsistent values to
remove from this variable and that there was previously no rejected
constraint. The trace also gives the inconsistent values that have
been withdrawn. Incidentally, the rule also specifies that the
constraint that prompted the reduction is still the active one ($A$ is
not modified).

Note that the upper part of the rules are not shown in actual traces,
this is {\em implicit} information. In most tracers, users have to
guess it. In our system, it is explicit, at least in the formal
specification.

\section{Implementation of a prototype tracer}
\label{sec:impl-prot-trac}

A formal specification of trace model is helpful to formally reason
about a trace. However, it actually specifies many traces for many
kinds of solvers. It is sometime too arid to decide whether possible
traces can be of any help to users.
With that respect, it is better to have samples of execution traces
checked by ``guinea pig users''. Producing samples by hand is very
tedious.
It becomes quickly  untractable since the trace model evolves
according to the remarks of the guinea pigs! New samples have
therefore to be produced until a trace model is validated by users.
It is, thus, welcome to have a tracer as quickly as possible.
However, a low-level implementation is not suited before the trace
model has been validated by users.
As already mentioned in the introduction, we propose to use a
prototype tracer in order to break this deadlock.

\subsection{Derivation of a prototype tracer from the CLP(FD) trace model}
\label{proto:section}

In our CLP(FD) experiment, we derive a CLP(FD) interpreter coded in Prolog
that we instrument with trace hooks.
Figure~\ref{meta:interpreter:figure} contains the translation of
\reject{} and \reduce{} rules of Figure~\ref{event:rules:figure}.
Each rule is encoded by a predicate with the same name as the rule.

\begin{figure}
\begin{listing}[1]{1}
reject(St0, St) :-
  St0 = ([C], S, Q, T,  [], D),
  get_varC(C, VarC),
  member(X, VarC),
  get_domain(X, D, []),
  St  = ( [], S, Q, T, [C], D),
  trace(reject, C, St0, _, _, _).

reduce(St0, St, ModOut) :-
  St0 = ([C], S, Q, T, [], D0),
  get_varC(C, VarC),
  member(X, VarC),
  reduction(C, D0, X, Wx),
  update_domain(X, Wx, D0, D, ModOut),
  St  = ([C], S, Q, T, [], D),
  trace(reduce, C, St0, X, Wx, ModOut).
\end{listing}
\caption{Translation into Prolog of the \reject{} and \reduce{} rules of Figure~\ref{event:rules:figure}}
\label{meta:interpreter:figure}
\end{figure}

Before paraphrasing the code for one rule, we give the meaning of all
the (simple) predicates that are used and not defined:
\Code{get\_varC(C, V)} takes as input a constraint \Code{C} and outputs
a list of the constraint variables that appear in \Code{C};\quad
\Code{member(X, L)} is a standard Prolog predicate which checks
whether \Code{X} is an element of the list \Code{L};\quad
\Code{get\_domain(X, D, Dx)} takes a constraint variable \Code{X} and a
domain state \Code{D}, and outputs the domain of $X$ (\Code{Dx});\quad
note that in Prolog \Code{=} denotes a unification;\quad
\Code{trace(Port, C, St0, Att1, Att2, Att3)} takes as input the
different event attributes; it calls the trace analysis system which
can, for example, print a trace line;\quad
\Code{reduction(C, D, X, Wx)} takes as input a constraint \Code{C}, a
domain state \Code{D}, a constraint variable \Code{X}, and outputs the values
to withdraw from the domain of \Code{X}; it fails if \Code{Wx} $= \emptyset$ (no
reduction can be done);\quad\\
\Code{update\_domain(X, Wx, D0, D, Mod)} takes as input a constraint
variable \Code{X}, a value set \Code{Wx} (to withdraw), and a domain
state \Code{D0}; it outputs the state domain \Code{D} such that
\Code{Dx=D0 - Wx}, and the list of modification types $X_{mod_1}$,
..., $X_{mod_n}$, where $mod_i \in \{min, max, ground, any,
empty\}$. These modifications characterize the \Code{Wx} value
withdrawal.

All the predicates implementing transition rules take as input a
solver state \Code{St0} and output a new solver state \Code{St}.
\Code{St0} and \Code{St} respectively denote the state of the solver
before and after the application of a rule.
%
%
Predicate \Code{reject(St0, St)} implements the \reject{} rule.  That
rule needs to fulfill 3 conditions to be allowed to be applied:
firstly, there is a constraint \Code{C} in the active constraint set,
secondly the set of rejected constraints is empty, and thirdly there
exists a variable attached to the \Code{C} constraint whose domain is
empty. This last condition implies the $unsatisfiable(c,{\cal
D})$ predicate of Figure~\ref{event:rules:figure}.
The first two conditions are checked line~2. The third condition is
checked lines 3 to 5 using Prolog backtracking. The execution of this
predicate will backtrack to \Code{member(X, VarC)} until either there
is no more variables in \Code{VarC} or one of them has an empty
domain.
After the application of the rule the set of active constraints is
empty and the set of rejected constraints contains $C$ (line 6). 
The tracer is called with the relevant information (line 7). Note the
use of Prolog anonymous variables for arguments 3 to 6. This means
that the values are meaningless.
The other rules are implemented in the same way.

The rules are integrated into the underlying Prolog system in the
usual meta-interpretation way of Prolog. This is not detailed in this
article. An introduction to meta-interpreters in Prolog can be found in
\cite{sterling94}.

\subsection{Discussion}

We have been able to systematically deduce the Prolog code from the
specification.
Even if the production of the prototype tracer was not
automatic, it has been sufficiently systematic so that the
modifications of the trace model could be immediately integrated.  
Furthermore, it has always been easy to convince ourselves that the
produced prototypes indeed implemented the expected trace model.\\

The produced prototype tracer was rather slow, but as already
mentioned in the introduction, efficiency is not an issue at this
stage. The idea is to validate the trace model, the tracer designers
use the prototype on well chosen and small programs.\\

The formal specification is very useful, but if time and resources are
short, and if the prototype tracer is very simple and easy to
implement, it can be considered to start the design of the tracer with
the prototype. One will miss the formal support for verification but
at least there will be some means to think about the trace design.

In our experiments, we have used the meta-interpretation capabilities
of logic programming to build  prototype tracers.
Even in languages with no meta-programming capabilities,
there are always some means to easily produce prototypes. For example
a source-level instrumentation of a subset of the language is
generally possible.
Not all the features and libraries of a language need to be
instrumented in the prototype. 
If the syntax of the language is specified by a grammar, then the
instrumentation can be implemented inside this grammar, for example
with  an attributed grammar language such as Yacc.\\

\section{Trace analysis}
\label{sec:trace-analysis}

Once a first trace model is designed and a prototype tracer exists, it
is important to play with actual traces. As soon as the traces are
longer than what can be printed on a page it becomes very tedious to
analyze them by hand. Even toy programs can produce traces of 
thousands of events whose systematic display would be inefficient and
irrelevant.  Therefore, we believe that the validation can be much
more rigorous and powerful if it is automated.  In our approach we
connect the prototype tracer with an automated trace analyzer \`a la
Opium \cite{ducasse99}.
The first advantage is that many long traces can be systematically
tested. A second advantage is that abstract and graphical views can be
automatically built \cite{ducasse99b,jahier00c}. In the case of CLP we
have built a number of graphical views which helped us select from the
potential interesting trace information the ones that were really
important and the ones that were not so crucial.
It should be noted that, applying the methodology, the trace model that
we designed varied quite a lot between the first design
\cite{langevine01b} and the current one \cite{langevine03}.

In the remaining of this section we first present the generic trace
querying facility. We then sketch one example of interesting
graphical view of CLP(FD) executions which has been built with the
trace analyzer.

\subsection{Analysis of the produced trace}
\label{analysis:section}

In our trace analysis scheme, users can formulate queries in order to
investigate an execution trace.  The queries are formulated in the Prolog
language extended by two primitives: \Code{fget/1} and\\
\Code{get\_attr/2}.
\Code{fget/1} searches  for a specific event forward in the
execution trace,
\Code{get\_attr/2} retrieves data about the current
  event.

\begin{figure}[t,t,t]
  \begin{center}
    \includegraphics[width=14cm]{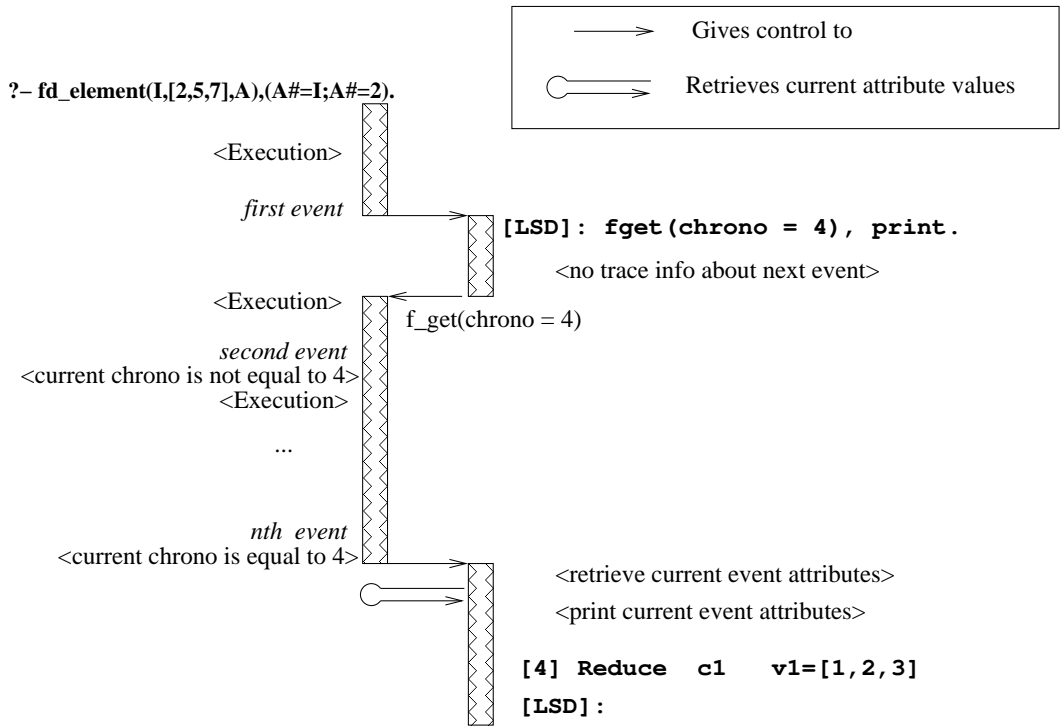}
    \caption{Illustration of the processing of a filtering query}
\label{synchro:filter:picture}
  \end{center}
\vspace{-0.4cm}
\end{figure}
 
Events are searched for as the traced program is executed.  There are
two execution processes, one for the traced execution, and one for the
trace analysis (called LSD in the following\footnote{LSD stands for a
``Long Story Debugger''. It is a prototype of generator of trace
analyzers currently under development.}).  
%
%
Figure~\ref{synchro:filter:picture} illustrates how the \Code{fget/1}
primitive works.  Let us assume that the programmer wants to query the
execution trace of the program given in
Figure~\ref{fig:trace:example}. When the execution reaches the
first event, it notifies LSD which prompts the programmer for a trace
query. The programmer enters a goal in order to search forward until
an event with chronological number equal to \Code{4} is found
(\Code{fget([chrono=4])}). This event should then be displayed
(\Code{print}).  At that moment, LSD can only get information about
the current event.  It therefore returns control to the traced
execution.  When the traced execution reaches the next event, it
locally checks whether the current chrono is equal to \Code{4}.  As
the current chrono is not the requested one, the traced execution is
resumed until the next event is reached. The chrono is again locally
checked. Forward moves and checking are done in turn until the first
event whose chrono is 4. LSD is notified and proceeds. The current
event attributes are retrieved by the \Code{print} command which
displays the related information. The execution of the trace query is
completed.  The programmer is then prompted for another one.

The scheme previously described is a good compromise between
efficiency and expressive power. On the one hand, the search for
events is done in the traced process, and can be very efficient. On
the other hand, as the whole power of Prolog is available in the
analyzer process, sophisticated debugging programs can be written.


\begin{figure}
\begin{listing}[1]{1}
:- fget([port = reduce, chrono>3]),
   get_attr([var, withdrawn], [X, Wx]).

:- setval(nb_reject, 0),
   fget(in(port, [reject, solution])),
   ( get_attr(port, reject)
     -> incval(nb_reject),
        fail
     ;  true                 
   ),
   getval(nb_reject, NbFailures),
   writeln(NbFailures).
\end{listing}
\caption{Two examples of trace queries}
\label{trace:query:examples}
\end{figure}

Two examples of composed queries are given in
Figure~\ref{trace:query:examples}.
The first query asks to go to the first event whose port
\footnote{following the Prolog tradition the type of events is called
``port''.} is \reduce{} with a chronological number ({\tt chrono})
greater than 3.
The reduced variable and the withdrawn domain are then retrieved and
stored in the variables \Code{X} and \Code{Wx}
On the trace given in Figure~\ref{fig:trace:example}, the query would
find the fourth event and return \Code{X=v1} and \Code{Wx=[0,4-268435455]}.

The second query of Figure~\ref{trace:query:examples} is an example of
sophisticated query which could be integrated into an analysis
program.  It counts the number of failures encountered before the
first solution and prints it.
First, the counter is initialized (line 4).
The \Code{fget} primitive is used to find the next event whose port is
either \reject{} or \solution{} (line 5). 
Then the actual port is retrieved with \Code{get\_attr} (line 6).
If it is \reject{}, then the counter is incremented (line 7) and a
failure is forced (line 8). 
If it is not a \reject{}, it means that it is a solution; in that case
the loop is stopped by simply executing \Code{true} (line 9).
In the case where the execution is made to fail, it will backtrack to
the \Code{fget} which will find the next event whose port is either
\reject{} or \solution{} (line 5). If the execution does no longer
contain such events, the overall query will simply fail.
In the case where the loop terminates on a \solution{} the value of
the counter is retrieved (line 11) and printed (line 12).

\subsection{A CLP(FD) visualization of variable updates}

\begin{figure}
\begin{center}
        \includegraphics[width=7.5cm]{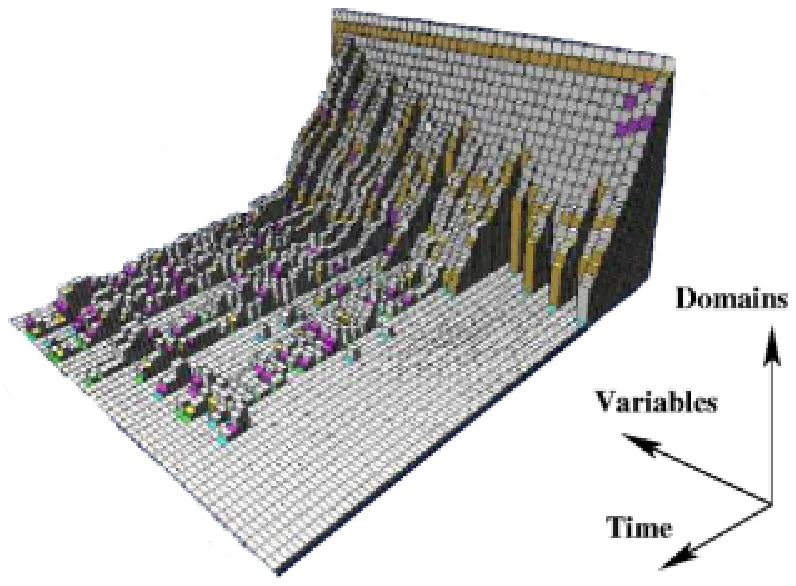}
        \hfill
        \includegraphics[width=7.5cm]{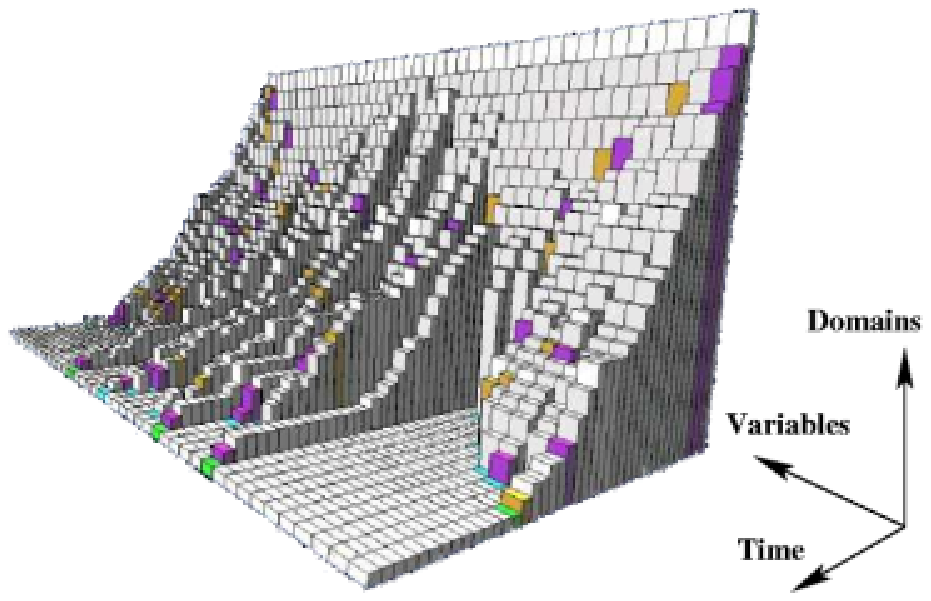}
\caption{Comparing two search procedures for the 40-queens problem with VRML
views computed by trace analysis.
}\label{FIG-VRML}
\end{center}
\end{figure}

One of our experiments is the generation of a 3D variable update view.
The evolution of the domains of the variables during the computation
is displayed in three dimensions. It gives a tool \emph{\`a la}
TRIFID~\cite{carro2000}.  The trace analyzer computes domain size each
time a constraint is added to the store or rejected, as well as when a
solution is found. The details of \reduce{} events allow us to assign
color to each kind of domain update (for example minimum or maximum
value removed or domain emptied) as made by Simonis and Aggoun in the
Cosytec Search-Tree Visualizer~\cite{simonis2000}.  The trace analysis
is implemented in about 125 lines of Prolog and generates an
intermediate file. A program implemented in 240 lines of C converts
this file into the VRML format.

Figure~\ref{FIG-VRML} shows the resolution of the 40-queens problem
with two different enumeration strategies. There are three axes:
variables, domain size and time. The first strategy is a first-fail
selection of the enumerated variable and the first value tried is the
minimum of its domain. The second strategy is also a first-fail
strategy but variable list is sorted with the middle variable first
and the middle of domain is preferred to its minimum.  The two
graphical views allow users to compare the efficiency of these two
strategies by manipulating the 3D-model. With the first strategy,
domain sizes on one side of the chess-board quickly decrease, and the
domain size on the other side oscillate at length. With the second
strategy, domain sizes decrease more regularly and more symmetrically,
the solution is found faster. In fact, the second strategy, which
consists in positioning the queens starting from the center of the
chess-board, benefits more from the symmetrical nature of the problem.

\subsection{Discussion}

The list of all the variables attached to the problem is a relevant
information because graphical views such as the 3D-model display all
the domains at a glance whereas the solver handles only a small subset
of variables at a time.  The above experiments made clear that the
tracer must be able, if requested, to provide the whole state at each
event, namely the domains and the constraints of the problem.
Furthermore, the list of variables involved in a given constraint is
not straightforward to get from the implementation. However, this kind
of information is also crucial to build some graphical views which are
helpful to users.
Even if producing these types of information requires some
implementation efforts and may have a cost in terms of performance, it
is worth producing them. This was not possible to decide by just
looking at the formal semantics.

\section{Efficient implementation}
\label{sec:effic-impl}

When the trace designer knows what in the expected trace model is
important for which debugging functions. The actual implementation can
start. The events of interest have to be located in the compiler or
compiled code. It is also necessary to specify how to get the
information related to each event. At this stage, the tracer
implementors can decide that a piece of information is too tricky or
too costly to produce. This happens all the time in tracers produced
without this methodology. The essential difference here is that the
tracer implementors know what is kept or not and why. This is
important for users.  For example, if a particular graphical view
requires some information which cannot be generated in a particular
implementation, users will know straightforwardly and will not
discover it the hard way.

\subsection{The Gnu-Prolog tracer implementation}
\label{implementation:section}

Encoding a trace model that is not derived from the actual
implementation of the solver is a delicate task. The correspondence
between a formalized event and the code of the solver is not obvious:
some events can be almost simultaneous, or a single event can be
performed in several points of the code.
For example, the trace model provides a unified view of the domain
reduction with the \reduce{} rule whereas there are several places
to instrument in the code. Domain reduction is a crucial point in a
constraint solver and the corresponding code is highly optimized. In
Gnu-Prolog, there are four different cases for the domain reduction,
depending on the way the values to withdraw have been computed.  The
tracer handles each case with its peculiarities in order to get a
single reduction event. Whatever domain reduction routine is used, the
trace event will be a \reduce{} with standard attributes.

Another issue is the ability to proceed through the whole sets of
constraints and variables, as well as to allocate them unique
identifiers. The solver only handles pointers on data
structures. During the execution, a given pointer can be used for
several constraints and variables.  Moreover, at a given moment, the
solver focuses on a small subset of entities.  Therefore the tracer
has to maintain its own data structures to reference all the pointers
the solver handles. When the solver creates a new variable or a new
constraint, the tracer references the pointer on this new entity in a
specific table. This table associates to this pointer a new integer
identifier and some debugging data that can be useful in the
sequel. When the solver deletes some constraints and variables, the
corresponding entries are removed from the tracer table. This table
can be used to search for an identifier knowing the pointer on an
entity or to search for a pointer knowing the identifier of an
entity. Both of these uses are made in logarithmic time. Another
possible use is retrieving the list of all the variables or the list
of all the constraints.

The methodology has led to a trace model that is far from the
implementation of Gnu-Prolog. The trace has needed the instrumentation
of critical points in the solver. Nevertheless, the implementation has
been possible and the final model has a clear semantics that is easy
to understand for a constraint programmer.
Moreover, the resulting tracer is efficient: the time
overhead while executing a program without any trace output is between
5 and 30 percents (less than 10\% in most of our benchmarks). While
producing a very detailed trace (with almost all the attributes the
model provides), the time ratio against an untraced execution is
between 3 and 7.4. 
These performances are comparable to other debuggers known to be
efficient enough, for example the Mercury tracer of Somogyi and
Henderson~\cite{somogyi99} or the ML tracer of Tolmach and
Appel~\cite{tolmach95}.

It is worth noticing that the tracer has been implemented in such a
way that only the part of the trace which is required by a specific
analysis is constructed: users pay only for what they need.

\subsection{Discussion}

Two other tracers exist for constraint programming. The first one is
the trace mechanism of the Ilog Solver platform. Ilog Solver is a C++
library for constraint solving. Some virtual trace functions are
called at some specific points of the solver. By default, those
functions do nothing.  The developer can redefine them to produce his
own trace. The parameters of the trace functions are the attributes of
the corresponding trace events. Taking the critical example of domain
reduction, we see that trace events are guided by the implementation:
there are two events by special case Ilog Solver handles: an event
\emph{before} and an event \emph{after}. Their attributes depend on
the case that is active. At the opposite, our model and implementation
provide a single unified event with all the data at a glance.

The second existing tracer is an experimental tracer for Sicstus
Prolog. Its trace model is dedicated to Sicstus implementation.  The
tracer is based on a complete storage of the trace and postmortem
investigation. When a specific data on an event is asked for, the
tracer has to traverse the trace backwards until the data is found or
is able to be recomputed. Most of our trace model could be produced
this way but it is not realistic for real-life executions.

%

\section{Validation }
\label{sec:validation}

Another important problem when building a tracer is to validate the
result. In particular, it is  important to be sure that the produced
trace is indeed the expected one. As opposed to the prototype, the
efficient implementation has no obvious relation to the formal
specification.

Our methodology proposes to further take benefits of the prototype
implementation in order to compare the traces produced by the actual
tracer with the trace produced by the prototype. This comparison is
not expected to be a bijection in the general case. Indeed, some
events may not be implemented (see the discussion above), some
information may not be available.  In addition, the actual tracer may
also produce larger traces for example for the parts of the language
and libraries that were not taken into account in the prototype. As a
consequence, at present, the comparison between the two types of
traces has to be done by hand.

\subsection{Validation of the Gnu-Prolog tracer implementation}
\label{validation:section}

\begin{figure}
Prototype trace:
\begin{listing}[1]{1}
newVariable   v1 =[1,2,3]
newVariable   v2 =[2,5,7]
\end{listing}

Gnu-Prolog trace:
\begin{listing}[1]{3}
newVariable   v1 =[0-268435455]
newVariable   v2 =[0-268435455]
newConstraint c1  fd_element([v1,[2,5,7],v2])
reduce   c1   v1 =[1,2,3]   W=[0,4-268435455]
reduce   c1   v2 =[2,5,7]   W=[0-1,3-4,6,8-268435455]
suspend  c1
\end{listing}
\caption{Portions of traces, with some attributes produced by the
  prototype tracer and Gnu-Prolog tracer for the introduction of new
  variables}
\label{trace:portions:figure}
\end{figure}

For our experiment we compared the traces of the executions of some
small programs produced by the prototype tracer and the Gnu-prolog
tracer. Figure~\ref{trace:portions:figure} shows portions of trace
related to the introduction of a new variable. 
In the prototype tracer, the variables are directly introduced with
their specified domain. 
In the Gnu-Prolog solver, every time a new variable is introduced, its
domain initially contains all the possible values and the following
execution steps use the regular reduction mechanism to reduce the
domain to the one which was declared. 
In our example, the two events of lines 1 and 2 produced by the
prototype tracer correspond to the events of lines 3 to 8 produced by
the Gnu-Prolog tracer. These events appear in
Figure~\ref{fig:trace:example} and have already been explained in
section~\ref{sec:trace-example}.

\subsection{Discussion}
Besides the identification of some minor implementation problems, this
analysis led us to refine the constraint identifiers to take into
account the fact that in Gnu-Prolog some built-in constraints are
split into several simpler constraints. The need for this refinement
could probably have been detected by other means, the systematic
comparison of the two types of traces, however, was a good support to
find problems in the low-level implementation, and this quite early in
the life time of the software.

\section{Conclusion}

In this article we have presented a methodology to rigorously design
and implement tracers in 5 steps: 1) design a formal specification of
the trace model, 2) derive a prototype tracer, 3) analyze the produced
traces, 4) implement an efficient tracer, 5) compare the traces
produced by the efficient implementation and the prototype.

The methodology has been used within the context of logic programming
where there is a strong background on semantics. We, however, believe
that the state transition approach can be applied to specify formal
trace models for other programming paradigms.

Even if we advocate to follow the complete methodology, some of the
steps can be useful without the others. For example, even without a
formal specification, starting with an easy to build and understand
prototype tracer is already a major improvement over starting directly
by the implementation of a low-level tracer.

We have shown how this methodology has been used to design and
implement a real tracer for CLP(FD) which is able to efficiently
generate information required to build interesting graphical views of
executions.
The trace model has been designed following mainly user's concerns
whereas usual tracers are designed following mainly implementation
concerns. The resulting tracer has performances comparable to
efficient tracers, therefore the methodology improves the quality of
the produced trace, and does not prevent efficiency.

\section*{Acknowledgments}
The authors would like to thank Rachid Zoumman who implemented the C
code which generates the VRML used in some analyses. They are also
grateful to the OADymPPaC partners for fruitful collaboration. In
particular, Narendra Jussien and Jean-Daniel Fekkete helped tune the
model.

\end{document}